\documentclass[pdflatex,sn-mathphys-num]{sn-jnl}

\usepackage{graphicx}%
\usepackage{multirow}%
\usepackage{amsmath,amssymb,amsfonts}%
\usepackage{amsthm}%
\usepackage{mathrsfs}%
\usepackage[title]{appendix}%
\usepackage{xcolor}%
\usepackage{textcomp}%
\usepackage{manyfoot}%
\usepackage{booktabs}%
\usepackage{algorithm}%
\usepackage{algorithmicx}%
\usepackage{algpseudocode}%
\usepackage{listings}%
\usepackage{xcolor}%

\theoremstyle{thmstyleone}%
\theoremstyle{thmstyletwo}%

\theoremstyle{thmstylethree}%

\begin{document}

\title{\boldmath A waveform and time digitization mainboard prototype for the hybrid digital optical module of TRIDENT neutrino experiment}


\author[1]{\fnm{Guangping} \sur{Zhang}}

\author*[1]{\fnm{Yong} \sur{Yang}}\email{yong.yang@sjtu.edu.cn}

\author[2,1]{\fnm{Donglian} \sur{Xu}}

\affil*[1]{\orgdiv{School of Physics and Astronomy}, \orgname{Shanghai Jiao Tong University, MOE Key Laboratory for Particle Astrophysics and Cosmology, Shanghai Key Laboratory for Particle Physics and Cosmology}, \orgaddress{\street{200 Dongchuan Road}, \city{Shanghai}, \postcode{200240},\country{China}}}

\affil[2]{\orgdiv{State Key Laboratory of Dark Matter Physics, Tsung-Dao Lee Institute}, \orgname{Shanghai Jiao Tong University}, \orgaddress{\city{Shanghai}, \postcode{201210},\country{China}}}


\abstract{

The TRIDENT (Tropical Deep-sea Neutrino Telescope) experiment is a
next-generation underwater neutrino observatory planned for deployment
in the West Pacific Ocean, designed to detect astrophysical neutrinos
through Cherenkov radiation. The full-scale detector will consist of
approximately 1000 vertical strings, each equipped with 20 hybrid
digital optical modules (hDOMs) containing both photomultiplier tubes
(PMTs) and silicon photomultipliers (SiPMs) for comprehensive light
detection. This paper presents a custom-designed digitization
mainboard prototype for the hDOM, featuring simultaneous
32-channel PMT waveform digitization at 125 MS/s using commercial
analog-to-digital converters and 56-channel high-precision time
measurement through field-programmable gate array -implemented
time-to-digital converters. The system demonstrates excellent
performance in single photoelectron (PE) detection with clear
pedestal separation, maintains linear response up to 240 PEs, and
achieves sub-nanosecond timing resolution for PMT or SiPM pulse
edges.
}


\maketitle

\section{Introduction}\label{sec1}
Astrophysical neutrinos are unique messengers to study some of the
long-standing problems of the universe such as the origins and
acceleration mechanism of cosmic rays. The TRopIcal DEep-sea Neutrino
Telescope (TRIDENT)~\cite{TRIDENT:2022hql}, planned to be constructed in the West Pacific Ocean,
will cover multi-cubic-kilometer of seawaters with around 1000
vertical strings, each around 700 meters long and spaced 70-100 meters
apart, adopting a Penrose tiling geometry. Each string will carry 20 hybrid digital optical modules
(hDOMs) to detect Cherenkov photons emitted from high-energy charged particles
produced in neutrino interactions. 

\begin{figure*}[!htbp]
  \centering \includegraphics[width=0.9\linewidth]{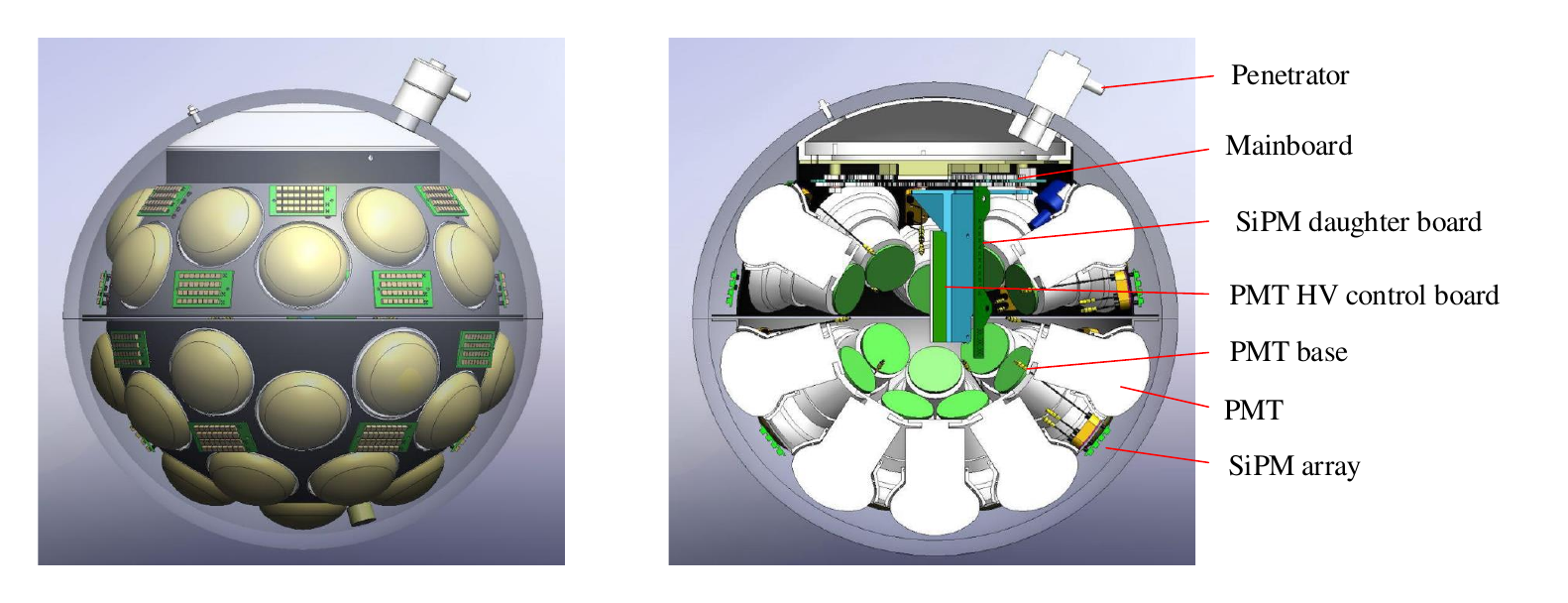}
  \caption{The overall and cross-sectional view of the TRIDENT prototype hDOM, including PMTs, SiPMs,
    the mainboard and other electronics boards. The hDOM is a pressure-resisted glass vessel with a diameter of 43 cm. }
  \label{hdom}
\end{figure*}

 Figure~\ref{hdom} shows the current prototype hDOM. Each hDOM will have a number of photomultiplier tubes (PMTs) for better photon
coverage. This also allows local coincidence to reduce backgrounds. Similar approach
is taken by the KM3NeT~\cite{KM3NeT:2019shc}, IceCube Upgrade~\cite{IceCube:2023upt}, and IceCube-Gen2 experiments~\cite{IceCube-Gen2:2023rji}.
TRIDENT will also use faster-response silicon photomultipliers
(SiPMs). The rising time of typical single photoelectron (SPE) signals
is about 1 ns~\cite{Zhi:2024dmr}. This will improve the photon arrival time measurement of each
hDOM and thus the pointing capability of the detector array~\cite{Zhi:2024dmr}. However, greater number of channels impose larger
challenges for electronics inside the module in terms of limited space (see Figure~\ref{hdom}),
power consumption budget, cost and so on. KM3NeT chooses to read out
the time-over-threshold (ToT) signals from PMTs and perform
digitization of these signals inside an FPGA to measure both the
arrival time and length of the ToT signals. For IceCube
Upgrade~\cite{IceCube:2023upt}, analog-to-digital converters (ADCs) with a sampling rate of 120 MS/s are used
for waveform digitization.  IceCube-Gen2~\cite{IceCube-Gen2:2023rji} takes a dual
readout approach; signals from each PMT are digitized with a 2-channel
ADCs at a sampling rate of 60 MS/s.

\begin{figure*}[!htbp]
  \centering \includegraphics[width=0.95\linewidth]{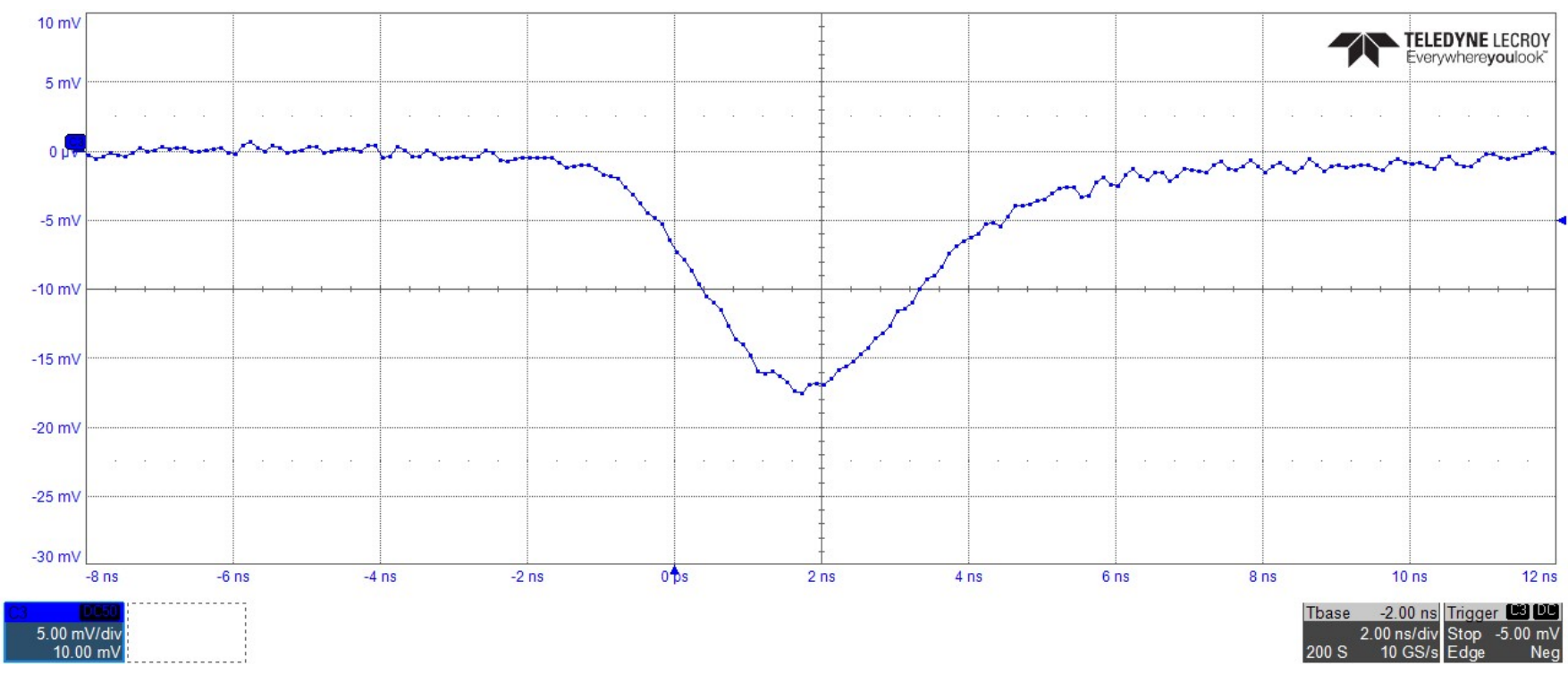}
  \caption{A typical SPE waveform from the PMT (Hamamatsu r14374)
    recorded by an oscilloscope with a 10 GS/s sampling rate.  The PMT
    is applied with a high voltage of 1200V.}
  \label{spe_lecroy}
\end{figure*}

The typical SPE signal from PMTs used by TRIDENT is
characterized by a short negative pulse with 2-3ns rising and 4-5ns
falling time (see Figure~\ref{spe_lecroy}). We would like to have the
digitized PMT waveform for possible pulse-shape analyses. A full
waveform also gives more precise measurement of photon counts than ToT
signals and opens new window for significantly improved efficiency for
astrophysical tau neutrino detection~\cite{Tian:2023+9,Tian:PhD}. Ideally, to preserve the original pulse shape, ADCs with a
sampling rate above several hundreds of MS/s are required. However,
high sampling rate usually means high power consumptions and
cost. Therefore, we aim to have a sampling rate around 100 MS/s. This
means we also need a pulse-shaping circuit in front of the ADC to
broaden the original PMT pulses. To mitigate the impact on the timing
measurements of relatively large sampling interval, the timing
information given by the ToT signals from the original pulses can be
measured more precisely using time-to-digital converter (TDC)
techniques.

In this paper, we present a digitization mainboard designed for
TRIDENT experiment, specifically engineered to support the first sea
trial of the hDOM prototype string scheduled for autumn 2025.  One
main functionality is that it can record both the digitized waveform
and rising time of signals from 32 PMTs. Its design and first
performances are presented.

\section{The Mainboard Design}
\label{design}

\begin{figure*}[!htbp]
  \centering \includegraphics[width=0.7\linewidth]{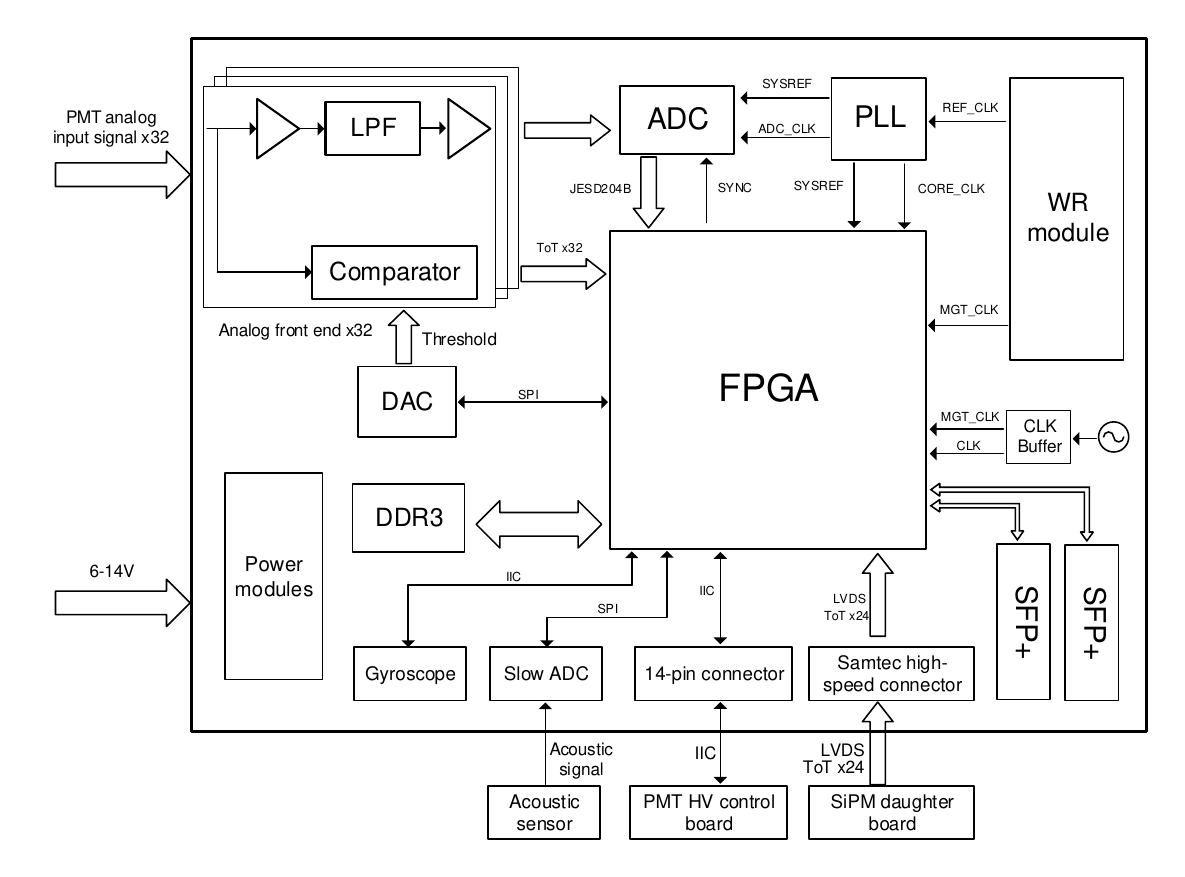}
  \caption{The overall design of the mainboard. Some of the main parts
    are: ADC (ADI AD9083), FPGA (Xilinx XC7k325T), DDR3 (Mircon MT41J128M16JT),  Comparator
    (TI LTV3603), and PLL (TI LMK04610).}
  \label{block}
\end{figure*}

\begin{figure*}[!htbp]
  \centering
  \includegraphics[width=0.45\linewidth]{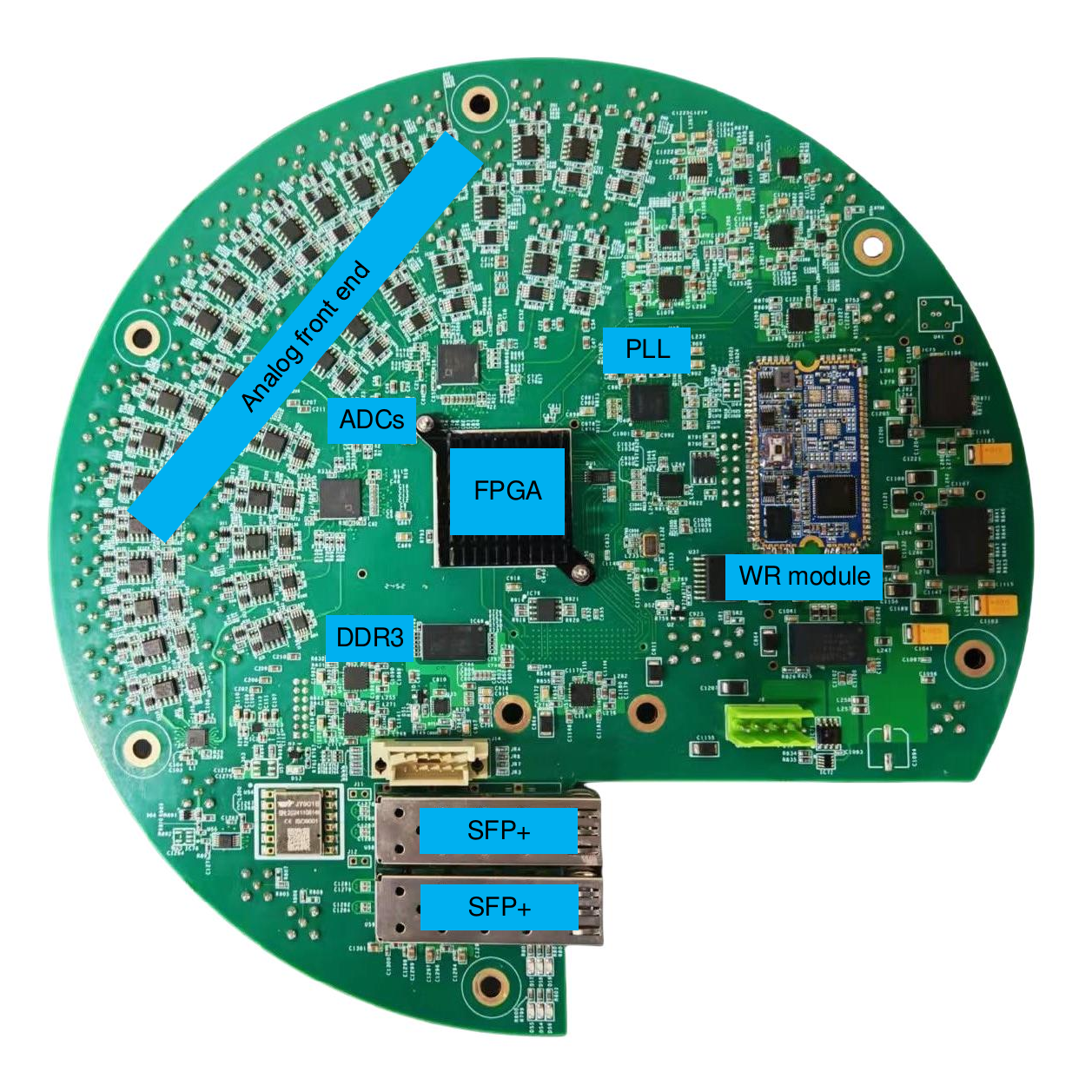}
  \includegraphics[width=0.45\linewidth]{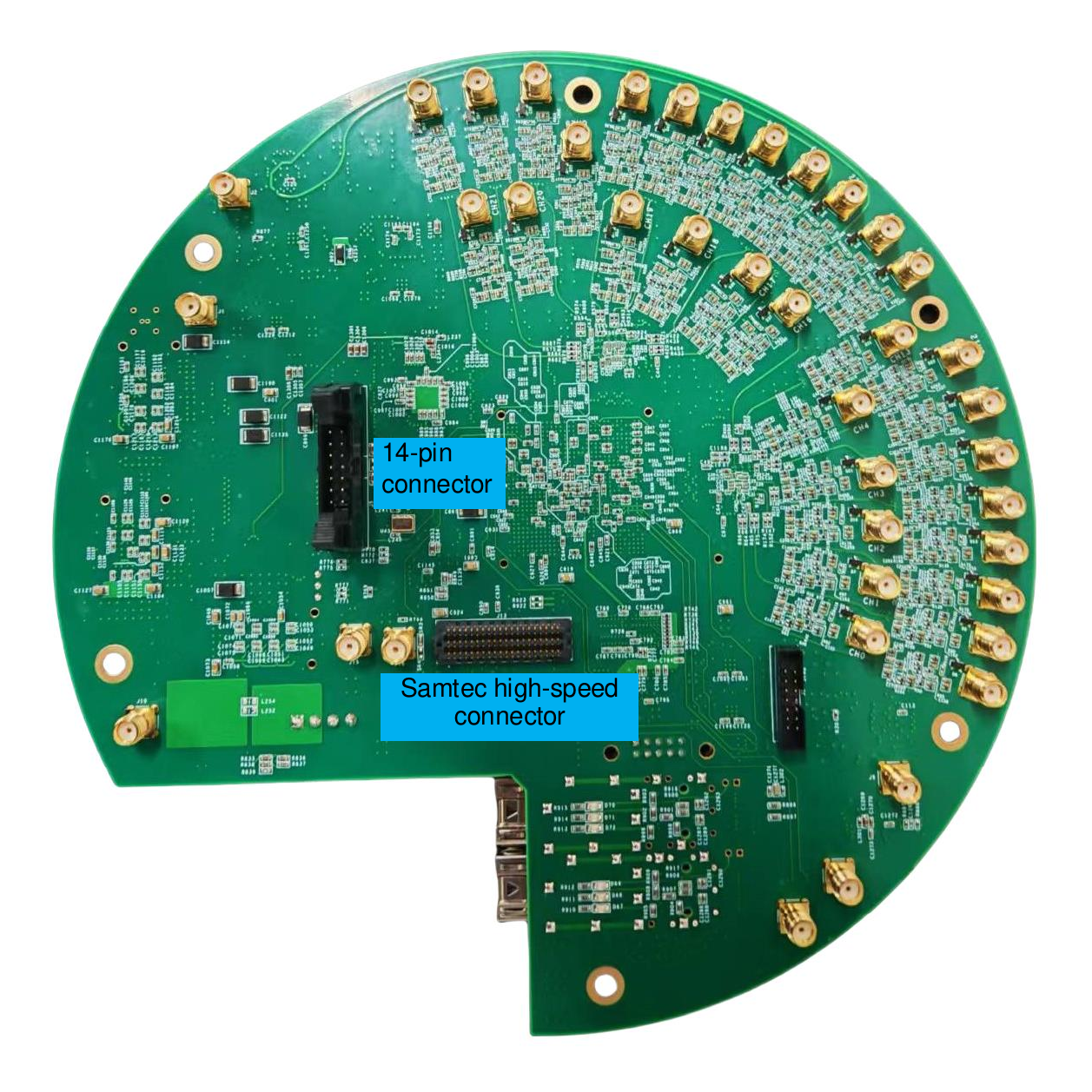}
  \caption{The top and bottom view of the digitization mainboard, which is a 16-layer PCB.}
  \label{board}
\end{figure*}

The design block of the mainboard is shown in Figure~\ref{block}. It can take at most 32 PMT
analog signals. All signal processing circuits for PMT
waveform digitization (ADC) and rising edge discrimination (TI LTV3603), digital
signal processing (FPGA, Xilinx XC7k325T), power and clock modules (TI LMK04610), and so on, are
integrated in one board. For this prototype, a 16-channel ADC from ADI (AD9083) is
chosen. It supports the JESD204B based high-speed serialized
output~\cite{jedec}, which helps to accommodate all components and
traces in a 22-cm-diameter PCB (Figure~\ref{board}). The ADC is configured, so each channel outputs 16-bit resolution digital data (input range of 2 Vpp) with
a sampling rate of 125 MS/s.

\begin{table}[!htbp]
  \centering
  \begin{tabular}{l p{6.5cm}}
    \toprule
    Parameter or fuctionality & Description \\ \midrule
    Number of ADC channels  & 32 instantaneous sampling at 125MS/s for PMT signals, and 1 channel at 1 MS/s for acoustic signal \\
    Number of TDC channels  & 32 for PMT and 24 for SiPM \\
    Clock synchronization   & WR integrated \\
    Trigger mode & Self-trigger, Coincidence trigger or External trigger \\
    PMT High voltage & Control and monitor  \\
    Accelerometer gyroscope & Monitor the angle, acceleration and so on \\
    \bottomrule
  \end{tabular}
  \caption{Key design parameters or functionalities of the mainboard.}
  \label{summary}
\end{table}

Table~\ref{summary} shows the key design parameters or functionalities
of the mainboard. In this paper, we focus on the waveform and
time digitization of the PMT signals.

\begin{figure*}[!htbp]
  \centering \includegraphics[width=0.9\linewidth]{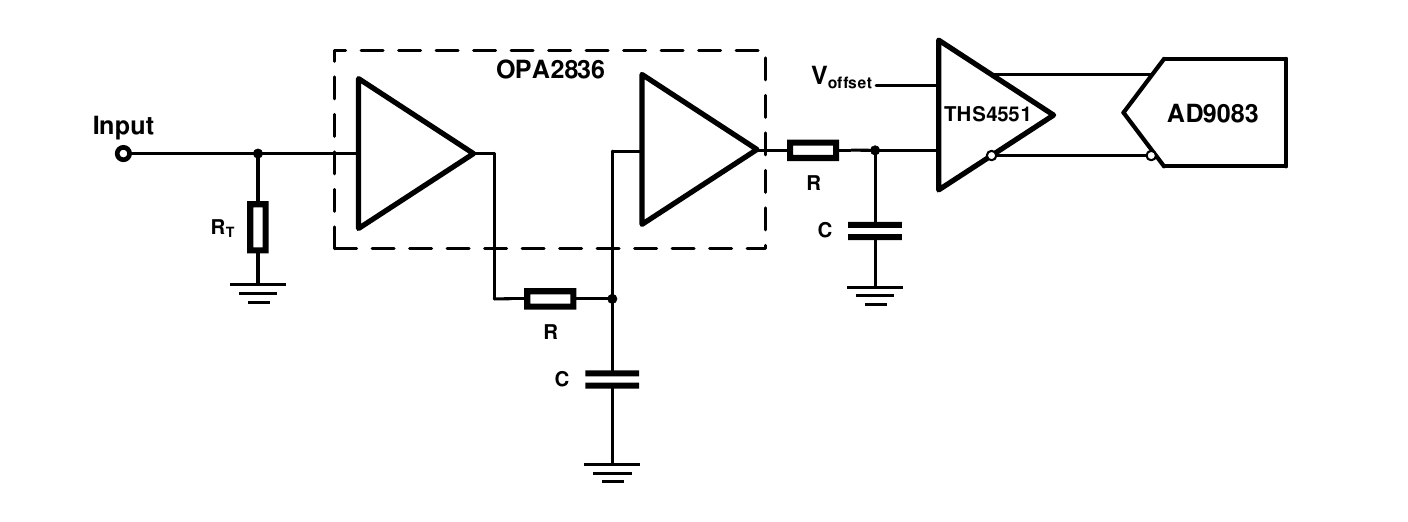}
  \caption{The signal conditioning circuit for ADCs. $R_{\mathrm{T}}$ is set to
    50 $\Omega$ for impedance matching, R$=50 \Omega$, C$=120$ pF for
    pulse broadening. $V_\mathrm{offset}$ is for baseline leveling.}
  \label{afe}
\end{figure*}

Each input PMT signal is processed by two analog front-end circuits in
parallel. One is the signal conditioning circuit for ADCs
(Figure~\ref{afe}), another one is the comparator circuit for TDCs.
In the first circuit, the single-ended input signal is first processed
with two preamplifiers and low pass RC filters. The first preamplifier
is mainly for impedance matching of 50 $\Omega$ and configured with unit
gain. But due to its limited bandwidth, it can introduce small
broadening effect. The two RC circuits perform the main broadening of
the pulse shape.  The second amplifier is set to be unit gain
(redundant design for this prototype).  Then it is converted into a
differential pair through a differential amplifier. The gain of this
amplifier is set to be two. Meanwhile, the baseline level is adjusted
to $\sim$0.9V to properly use the $\pm1$V dynamic range of the ADC. To
illustrate the effect of this circuit, for SPE signal, the rising edge
and falling edge are enlarged to about 15 ns and 30 ns,
respectively. The amplitude is reduced to approximately 7 mV. The
digital data from ADC are transferred to the FPGA for online
processing.

The mainboard integrates a custom-designed White-Rabbit (WR)-based
timing module from Sync(Beijing) Technology. This WR module can provide
a synchronized clock with sub-ns precision when the mainboard is connected to
a upstream WR switch through a small form-factor pluggable plus transceiver (SFP+) port with
a fiber optical link. This clock is used as the
reference clock for a JESD204B compliant and programmable clock jitter
cleaner with internal phased-locked loops (PLL) on the mainboard. This
clock cleaner generates necessary clocks for ADC and JESD204B receiver
core clock in the FPGA, as well the system reference (SYSREF) signal
to achieve deterministic latency among multiple ADCs (the JESD204B
subclass 1 system).  In addition, several other clocks are
required. One clock is needed for configuration (e.g. the PLL and
ADCs).  Serial transceivers in the FPGA also require reference clocks
(denoted as MGT$\_$CLK in Figure~\ref{block}).  One MGT$\_$CLK is from
the WR module. Other clocks come from an oscillator on the board.

Inside the FPGA, data from ADCs are separated into
channel-by-channel. The way the data are recorded depends on the
configured trigger mode. For the self-trigger mode, each channel can
be read out independently with baseline suppressed when the incoming data exceed
the baseline by a configurable threshold. This mode can be used to record SPE signals.
For coincidence trigger mode, a minimum number of channels are required to
be simultaneously triggered within a time window. This mode is
designed to record physics events while rejecting dark-current
signals. For the external trigger mode, all channels of a fixed length
are read out upon receiving an external trigger request.  This mode is
useful for light calibration of PMTs. To record waveform before the
trigger, first-in-first-out (FIFOs) in the FPGA are used to cache the
data. Recorded ADC data and other information such as the trigger time
stamp and channel number are written into separate FIFOs before they
are buffered into the DDR3 (Mircon MT41J128M16JT).

In this second circuit, each PMT input signal is fed into a comparator
which generates a ToT signal. The threshold of each comparator can be
independently adjusted by the output of a DAC.  The rising time of
each ToT signal is measured by a tapped delay line-based TDC
implemented inside the FPGA, using the CARRY4 block, each of which has 4 delay taps.
In this work, each TDC line consists of 96 CARRY4s, spanning two clock regions.
Each measurement consists of a coarse counter (with TDC system clock of 4 ns period) and a fine counter
which refers to the number of taps delay between the rising edge of
the ToT signal and the next nearest rising edge of the system
clock.  One unit of fine counter corresponds to 12 ps on average, which
limits the intrinsic precision of the TDC in FPGA. Similarly, the TDC data can be recorded depending on the
trigger mode. Together with other information, these data are written
into FIFOs and then to the DDR3. A Gigabit Ethernet protocol
SiTCP~\cite{4545224} is used to transfer these data from the FPGA to the
DAQ server through another SFP+.  In future, it might be possible
to use the same WR link to transmit data to reduce the number of
optical fibers along each string. This is still under development.

Besides the above-mentioned functionalities, the mainboard is also
designed to take at most 24 LVDS ToT signals from the Samtec
high-speed connector shown in Fig.~\ref{board}. These signals come
from another daughter card where each SiPM analog output is
discriminated against a threshold configured by the mainboard. Therefore, we can measure
the ToT signals from SiPMs using the same TDC techniques in the
FPGA. Another 1 MS/s ADC (slow ADC in Figure~\ref{block}) is used for
digitizing the acoustic signal. An accelerometer gyroscope is used to
monitor the angle, acceleration, angular velocity, magnetic field and
temperature.  Finally, this board has a 14-pin connector, which can be
used to control and monitor the HV of each PMT through a separate HV
control board.

The mainboard requires a DC power supply with an input range of
6-14 V. The input voltage is subsequently stepped down to various lower
voltages through power modules (DC-DC converters or linear dropout
regulators). Under the nominal condition of 12 V, the board draws
approximately 2.3 A. Although this power profile imposes no
constraints for the imminent hDOM prototype sea trial, optimization of power
consumption remains a priority for future iterations.

\section{Performance}

\begin{figure*}[!htbp]
  \centering \includegraphics[width=0.5\linewidth]{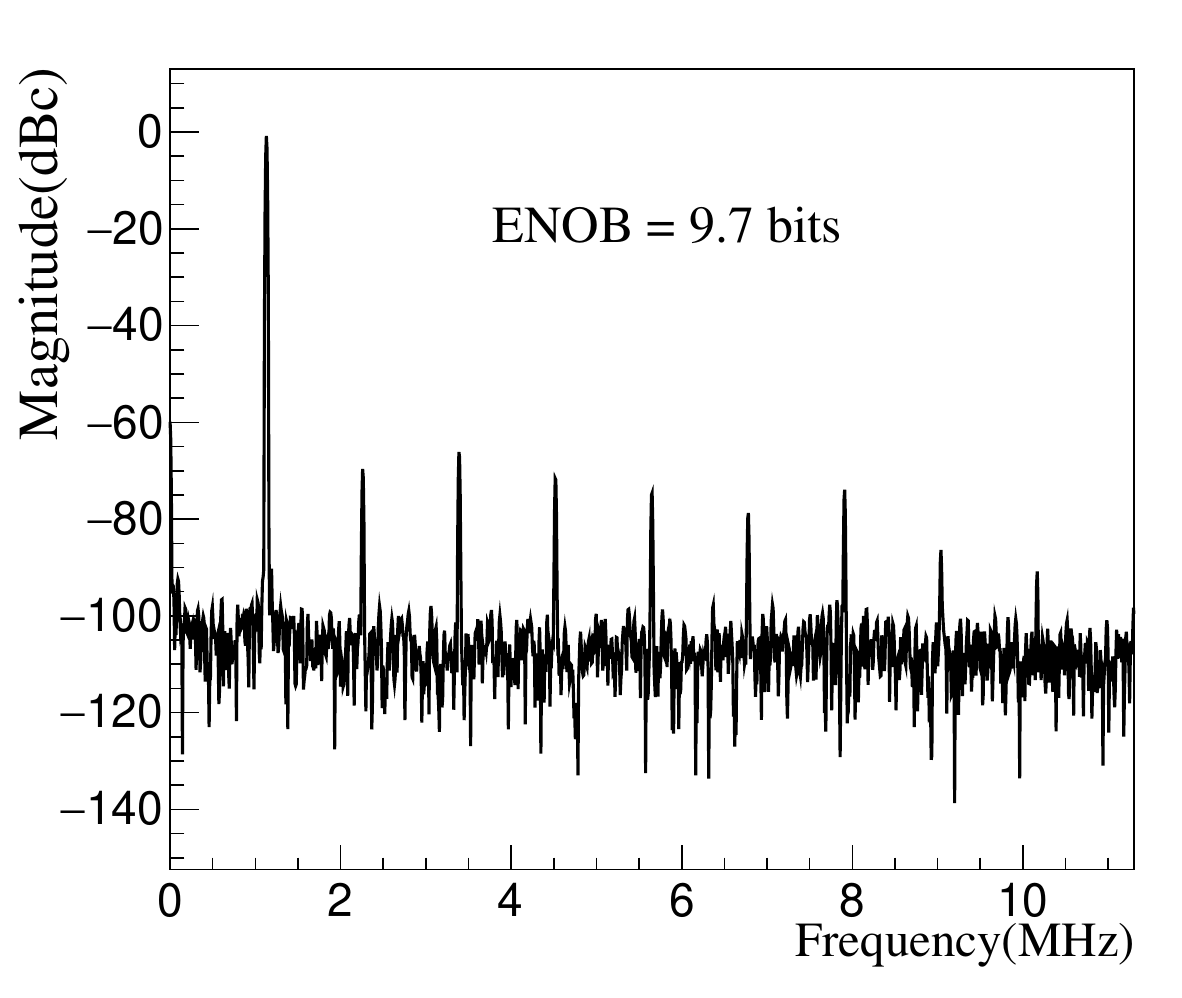}
  \caption{Frequency domain representation of the digitized samples
    with 1.13 MHz sin-wave input.}
  \label{fft}
\end{figure*}

In this design, ADC plays an important role. But it should be
emphasized any ADC performance might be affected by the front-end
amplifier circuits. Particularly, in our case, we have three
amplifiers. For dynamic performance, one major specification is the
effective number of bits (ENOB). To illustrate the performance, we use a 1.13
MHz sin-wave ($\sim0.9$ full scale) as input. An FFT analysis
(Figure~\ref{fft}) shows the ENOB is 9.7. Major static performance
parameters include differential nonlinearity (DNL) and integrated
nonlinearity (INL). However, for the chosen AD9083 ADC, these
parameters are hard to interpret due to its first order sigma-delta
architecture~\cite{forum}. 

In this paper, we focus on the experimental performance
of the mainboard related to PMT signals. We report the performance of the charge measurement
from the digitized waveform using ADCs, including the SPE signal and
the linear range, which are among the most important system-level
criteria for waveform digitization.  Afterward, we report the
resolution of the SPE rising-edge time measurement using the TDC
implemented in the FPGA.

\subsection{Charge Measurement}

\begin{figure*}[!htbp]
  \centering
  \includegraphics[width=0.9\linewidth]{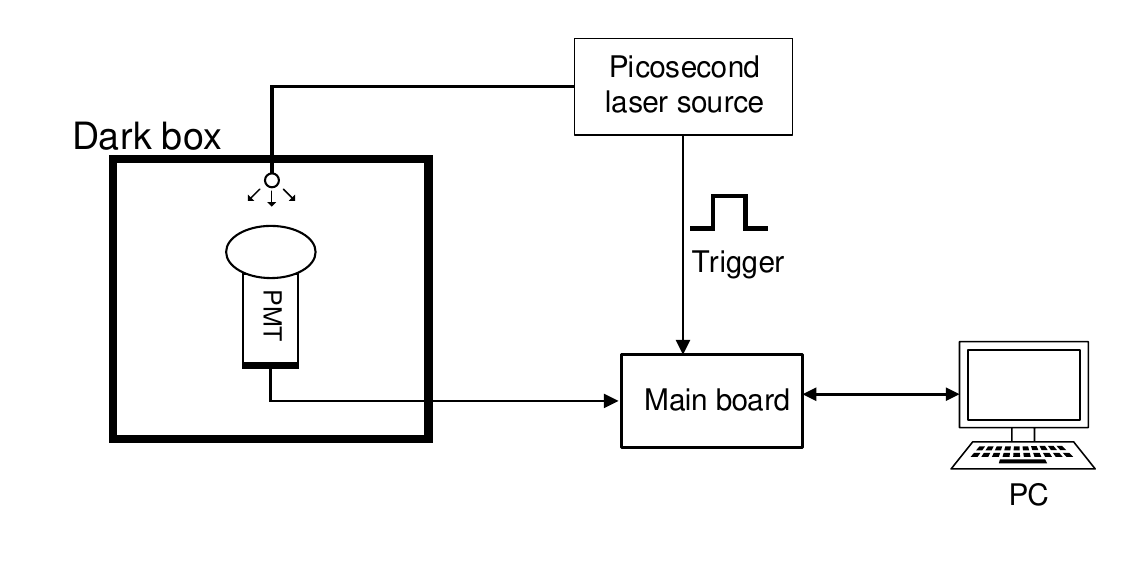}
  \caption{The setup for PMT charge measurement with the mainboard and a laser source.}
  \label{setup}
\end{figure*}

\begin{figure*}[!htbp]
  \centering \includegraphics[width=0.45\linewidth]{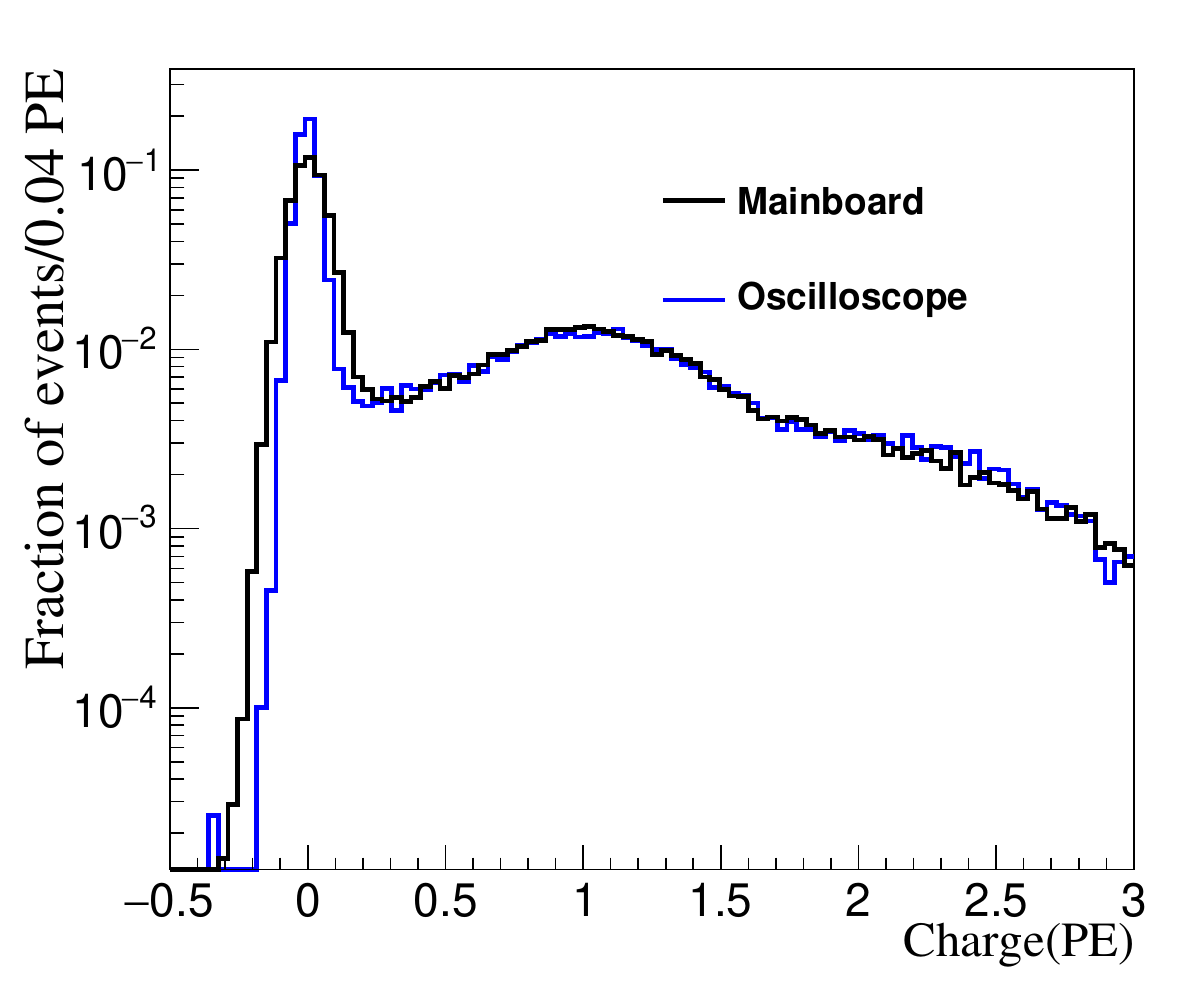}
  \includegraphics[width=0.45\linewidth]{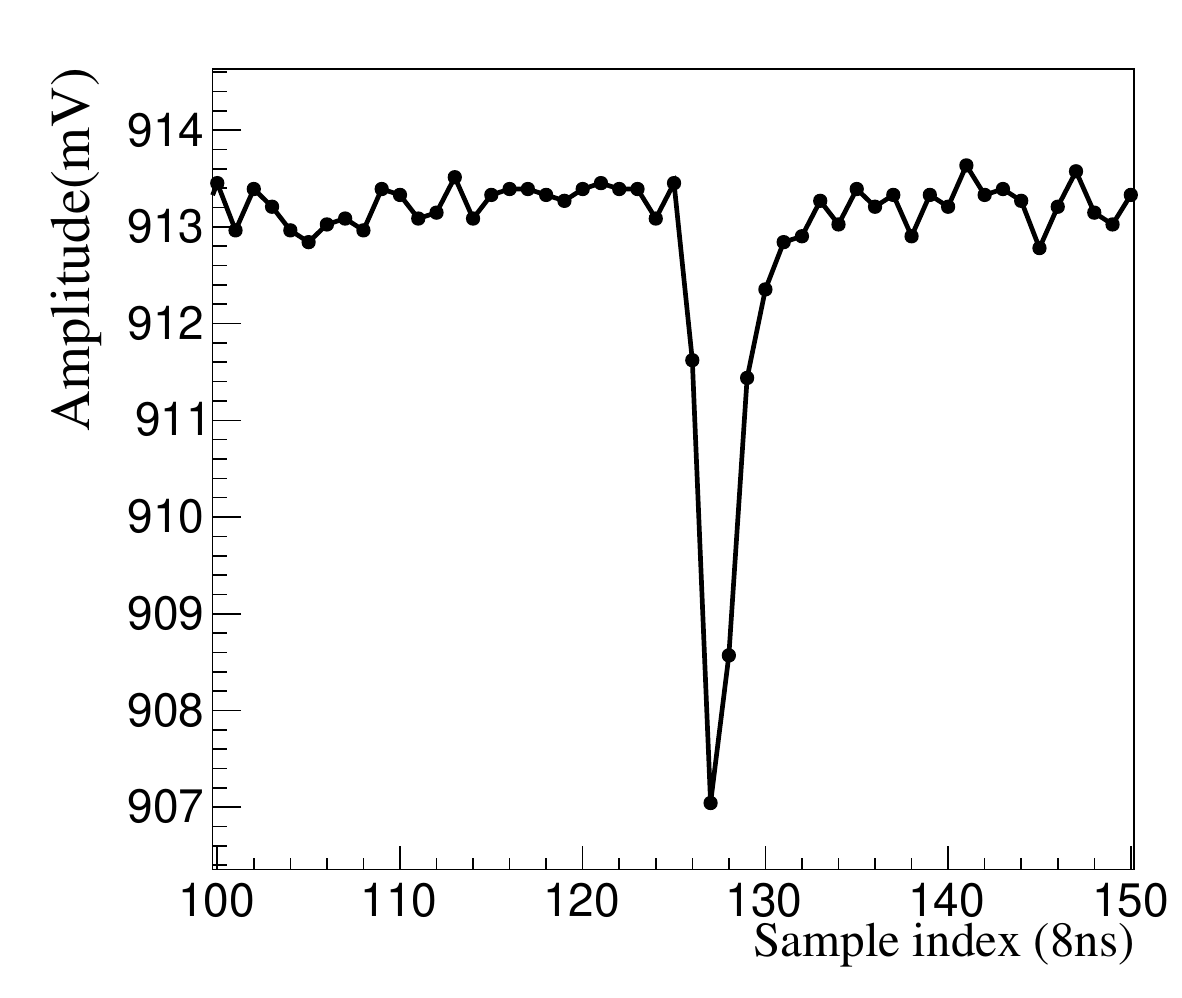}
  \caption{Left, the measured charge distribution with a low-intensity
    light source. Right, a typical SPE waveform recorded by the
    mainboard.}
  \label{spe}
\end{figure*}

A laser source is used to evaluate the performance of the waveform
digitization, shown in Figure~\ref{setup}. Pulsed laser lights (typical width of 10 ps) are
injected into the PMT together with a trigger request into the
mainboard. The light intensity is controlled with optical filters.
At each intensity, we measure the charge from a larger number of
events. We first tune the intensity for SPE calibration.
The left panel of Figure~\ref{spe} shows the measured charge distribution by
integrating the digitized waveform in a fixed time interval around the
pulse. The pedestal peak exhibits an RMS of 0.06 PE, coming from the
integrated electronic noise.  In contrast, the SPE signal peak
shows an RMS of 0.4 PE, dominated by the statistical
fluctuations in the PMT gain. This indicates that the noise
contribution becomes negligible for charge measurements of the PMT's
smallest signals, as the SPE resolution is overwhelmingly determined
by gain variation. At the same intensity, the PMT pulses are recorded
by an oscilloscope with a sampling rate of 10 GS/s. The noise performance is
much better in the oscilloscope, however, the SPE peak is almost the same
compared to the mainboard.

\begin{figure*}[!htbp]
  \centering \includegraphics[width=0.45\linewidth]{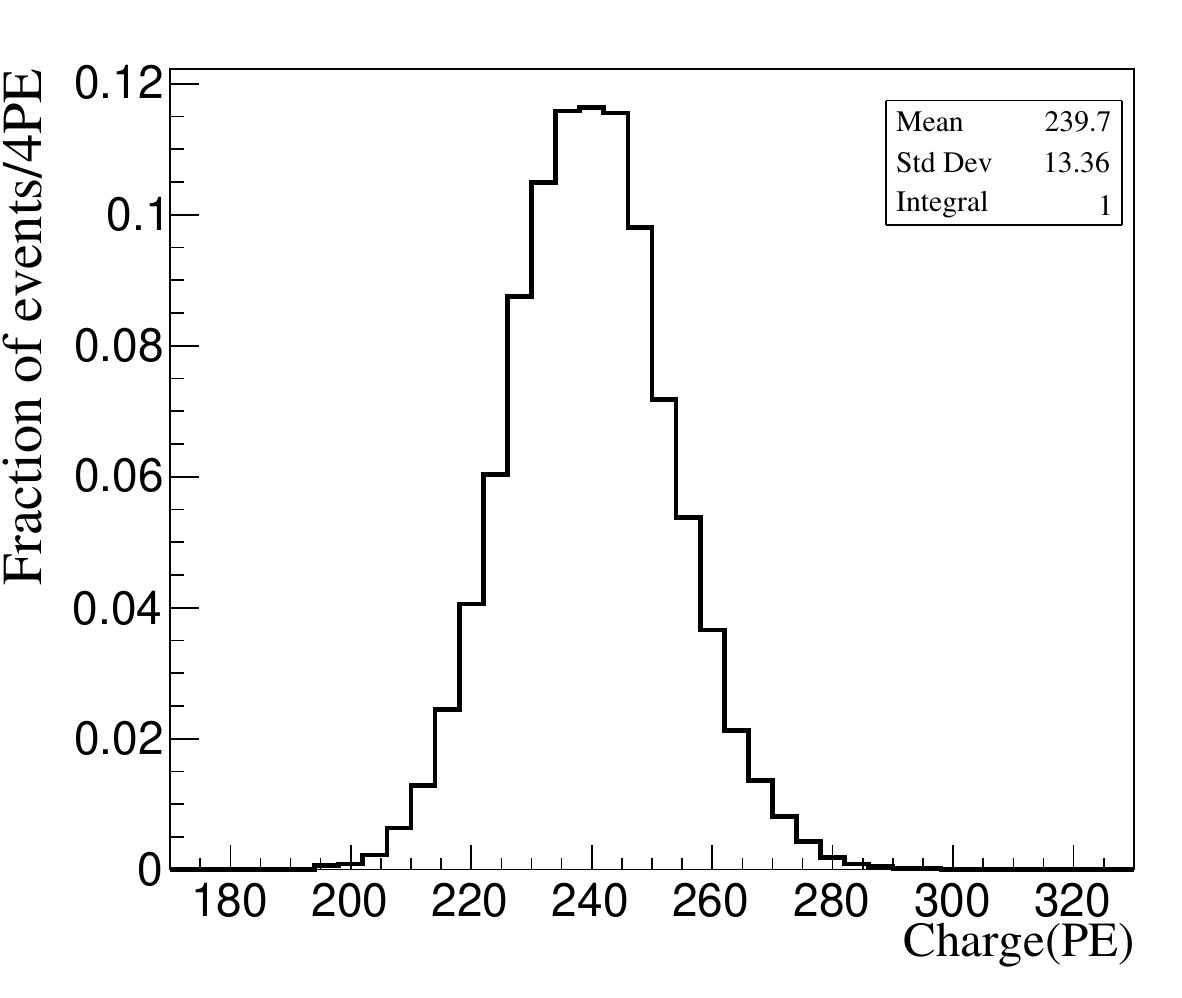}
  \includegraphics[width=0.45\linewidth]{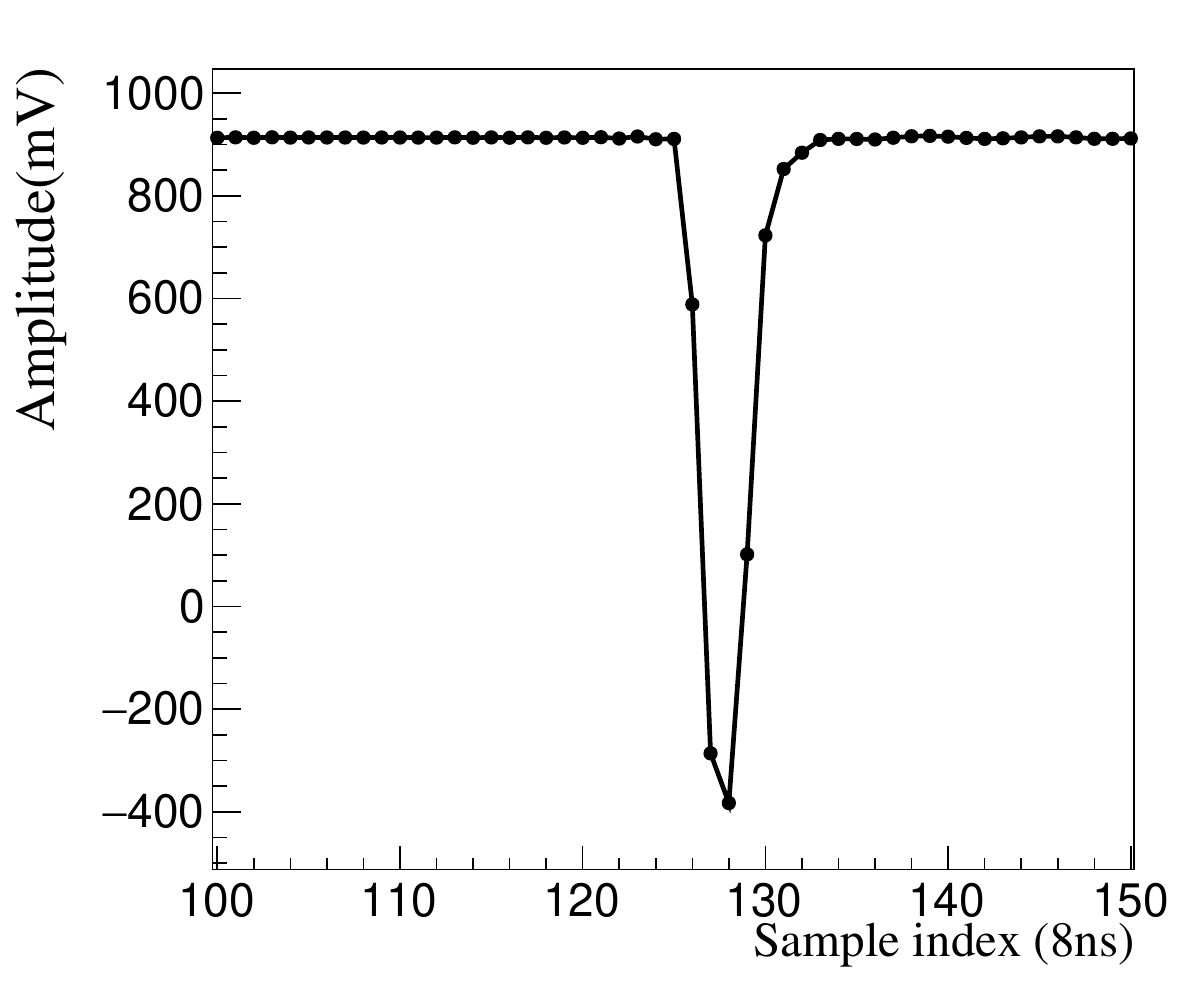}
  \caption{Left, the measured charge distribution with a
    high-intensity light source. Right, a typical waveform recorded at this light intensity.}
  \label{highcharge}
\end{figure*}

The right panel of Figure~\ref{spe} shows a typical waveform of
the SPE signals recorded by the mainboard. The intensity of the laser
source is then tuned. Figure~\ref{highcharge} shows the measured
charged distribution and a typical waveform at the largest intensity
of the laser source.

\begin{figure*}[!htbp]
  \centering \includegraphics[width=0.5\linewidth]{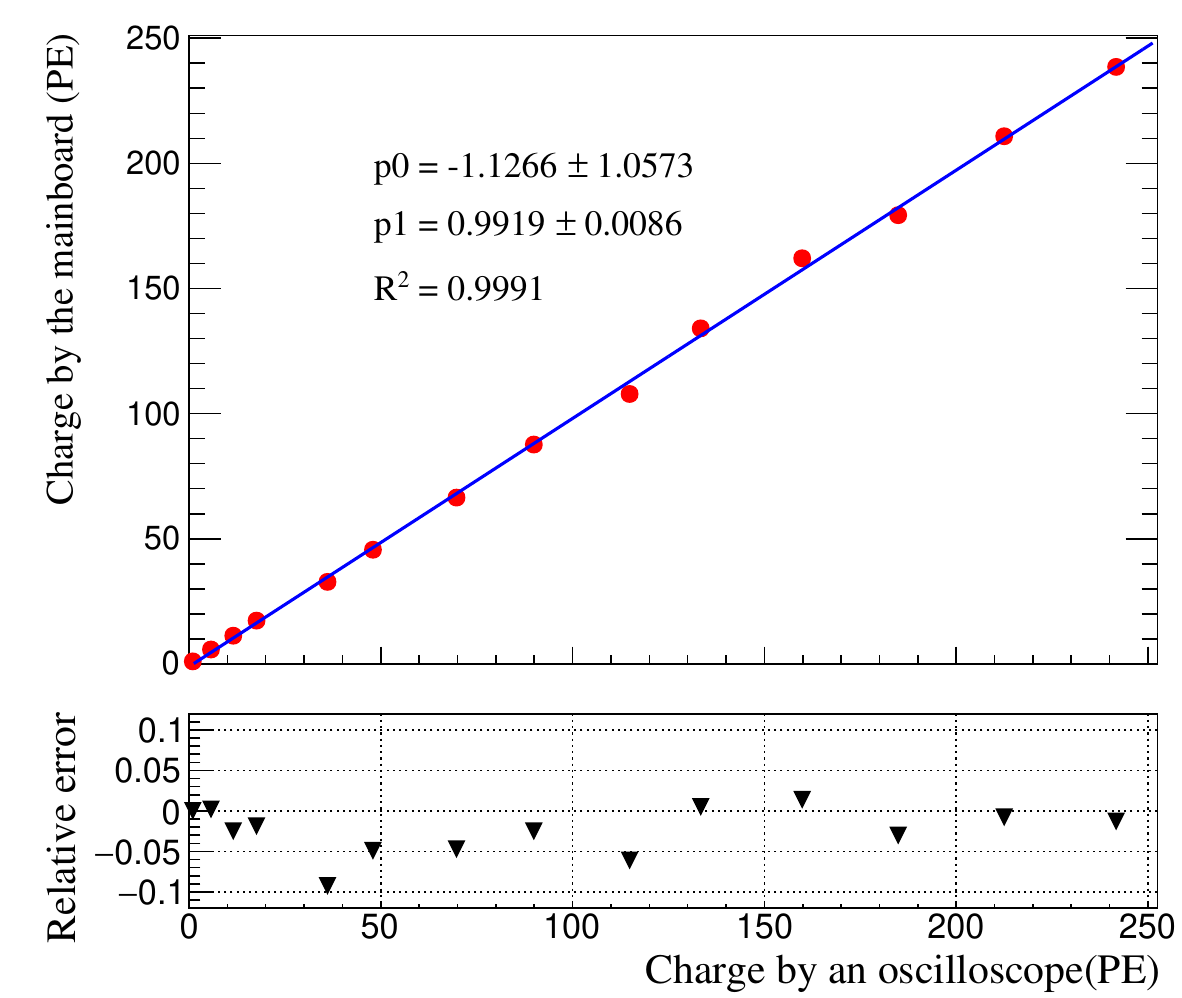}
  \caption{The linearity test result of the mainboard with a PMT and a
    light source. Reference measurement is from an oscilloscope. The data are fitted with a linear curve
    $y=p0 + p1\times x$. }
  \label{linearity}
\end{figure*}

To evaluate the linearity performance of the mainboard, we compare the
charge measured from the mainboard with that from the oscilloscope,
because the intensity of laser source is not
calibrated. Figure~\ref{linearity} shows that our mainboard has good
linearity up to 240 PE.  The data are fitted with a straight line, the
obtained R-squared value is 0.999. At all light intensities, the relative deviations
of the mainboard's measurements compared to the oscilloscope are within 10\%. Overall, the mainboard's results
are slightly lower. For comparison, IceCube collaboration~\cite{IceCube:2023upt} reported a linear dynamic range
of 150 PE with their DOM mainboard for the Upgrade
experiment. IceCube-Gen2 collaboration~\cite{IceCube-Gen2:2023rji}
uses dual-readout design for each PMT (high gain with anode and low
gain with eighth dynode), achieving a linear dynamic range of 40 PE
(high gain) and 2500 PE (low gain).

\subsection{Timing measurement}

\begin{figure*}[!htbp]
  \centering \includegraphics[width=0.45\linewidth]{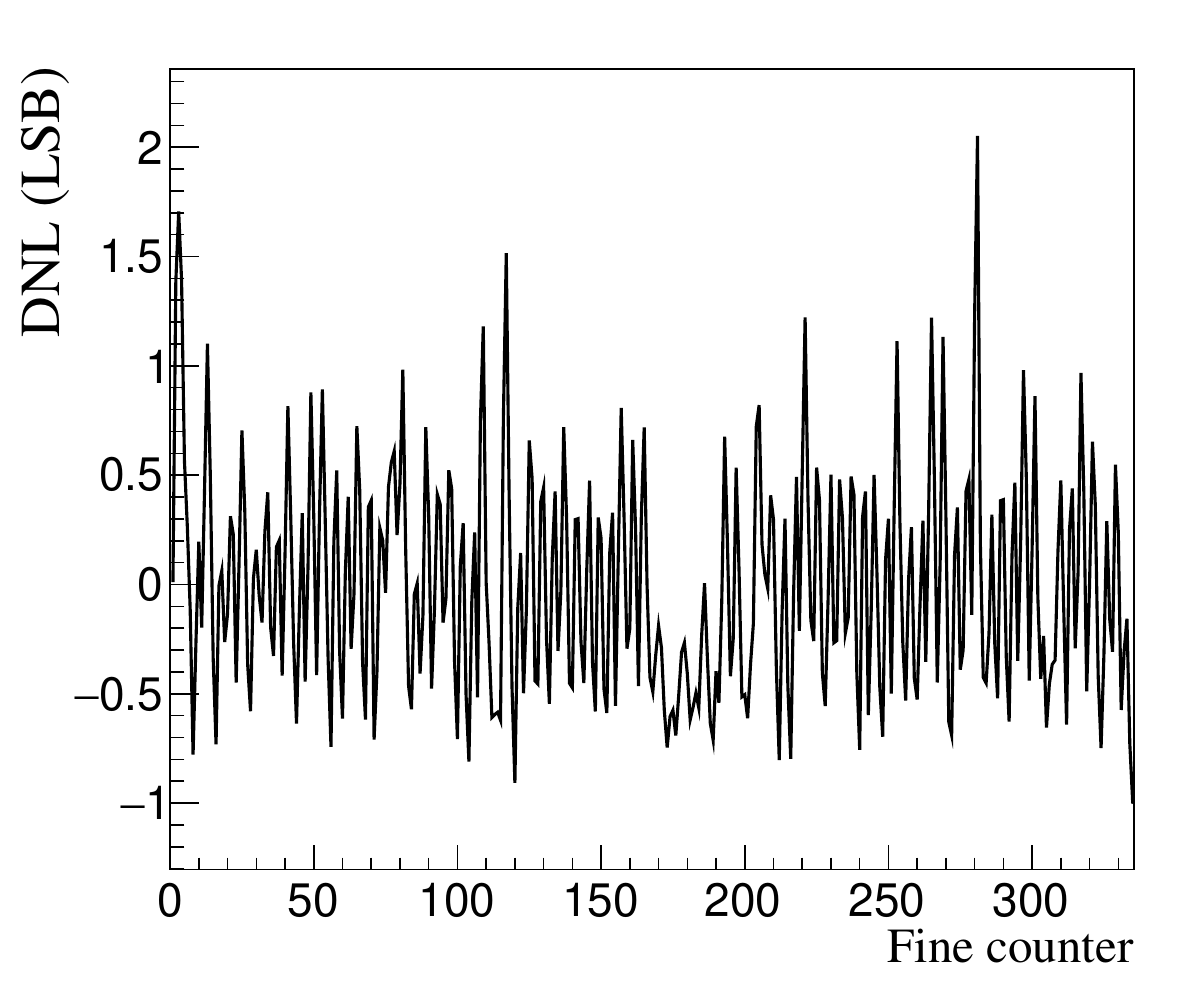}
  \centering \includegraphics[width=0.45\linewidth]{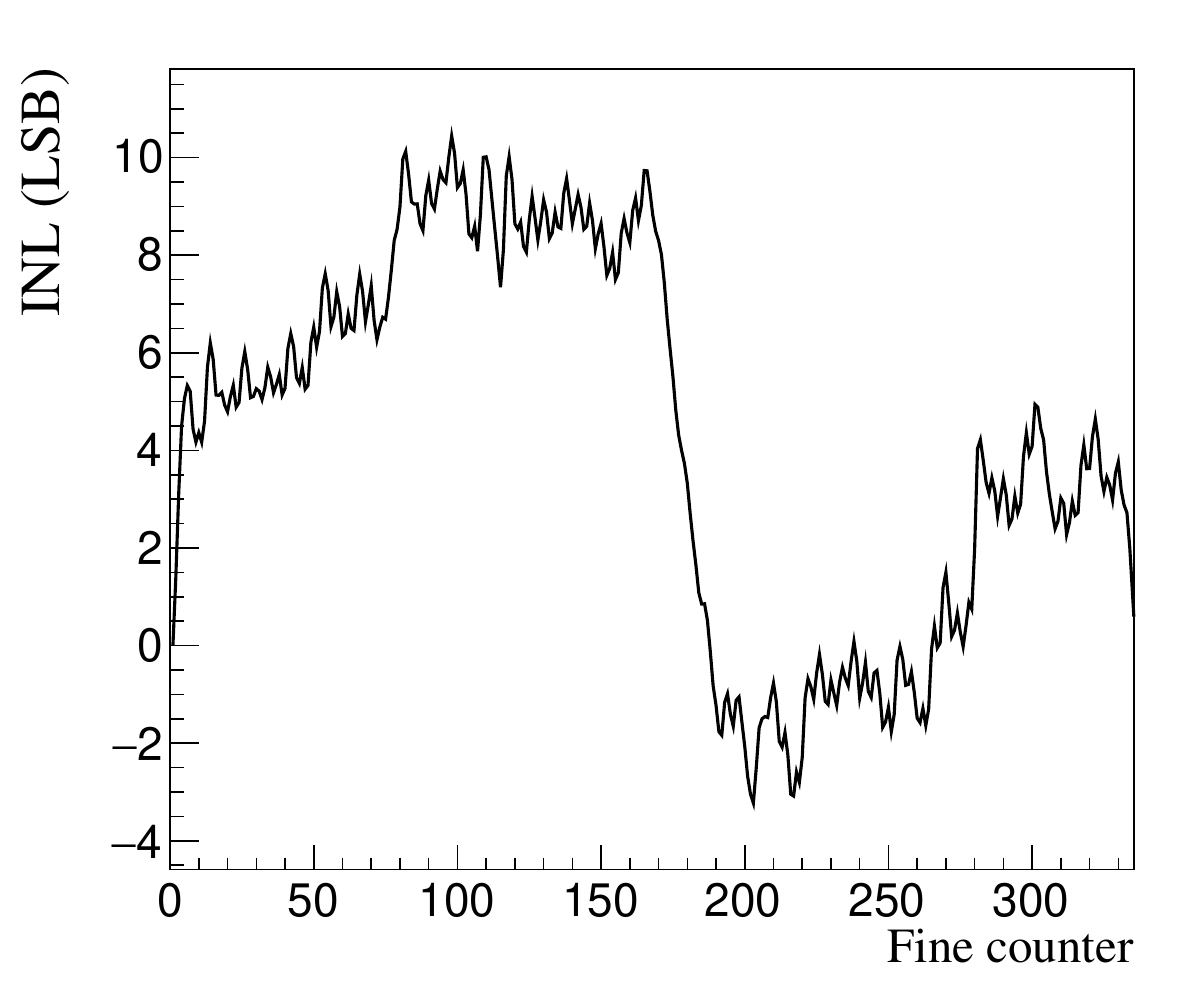}
  \caption{The DNL (left) and INL (right) of the TDC. The LSB refers
    to the average delay of 12 ps. The feature around fine counter of
    180 is caused by the two clock region boundaries in the TDC delay
    line.}
  \label{dnl}
\end{figure*}

As mentioned above, we also want to measure the rising edge time of the ToT
signal of each PMT channel with the delay chains inside the FPGA. For each TDC
channel in the FPGA, the delay of each element along the delay line is
not identical, but can be measured from the
distribution of the fine counter by collecting a large amount of
events which are asynchronized with the TDC system clock. This is the
so-called bin-by-bin calibration or statistical code density test. We
can use SPE events to perform this calibration {\it in-situ}. The
average delay is measured to be 12 ps (denoted as LSB of the TDC).
As shown in Figure~\ref{dnl}, the relative difference of each delay to the average, namely the
DNL, ranges from -1 LSB to 2 LSB, and the INL ranges from -3 LSB to 10 LSB. To estimate the
intrinsic time resolution of the TDC, two identical digital pulses are directly sent to
two TDC delay lines via two SMA connectors on board. The RMS of the
time difference is about 12 ps.

In addition, signal propagation delay from the input port to the first element in
the FPGA varies among different channels. These channel-by-channel
variations can be calibrated using events with the same arrival
time.  In the laboratory, we can use a signal-splitter board to
distribute the same pulse with equal-length cables to 32 input
ports. {\it In situ}, this can be calibrated using the Potassium-40
backgrounds from the seawater. This {\it in situ} calibration can also
remove possible systematic transit-time variations among different PMTs.

\begin{figure*}[!htbp]
  \centering \includegraphics[width=0.5\linewidth]{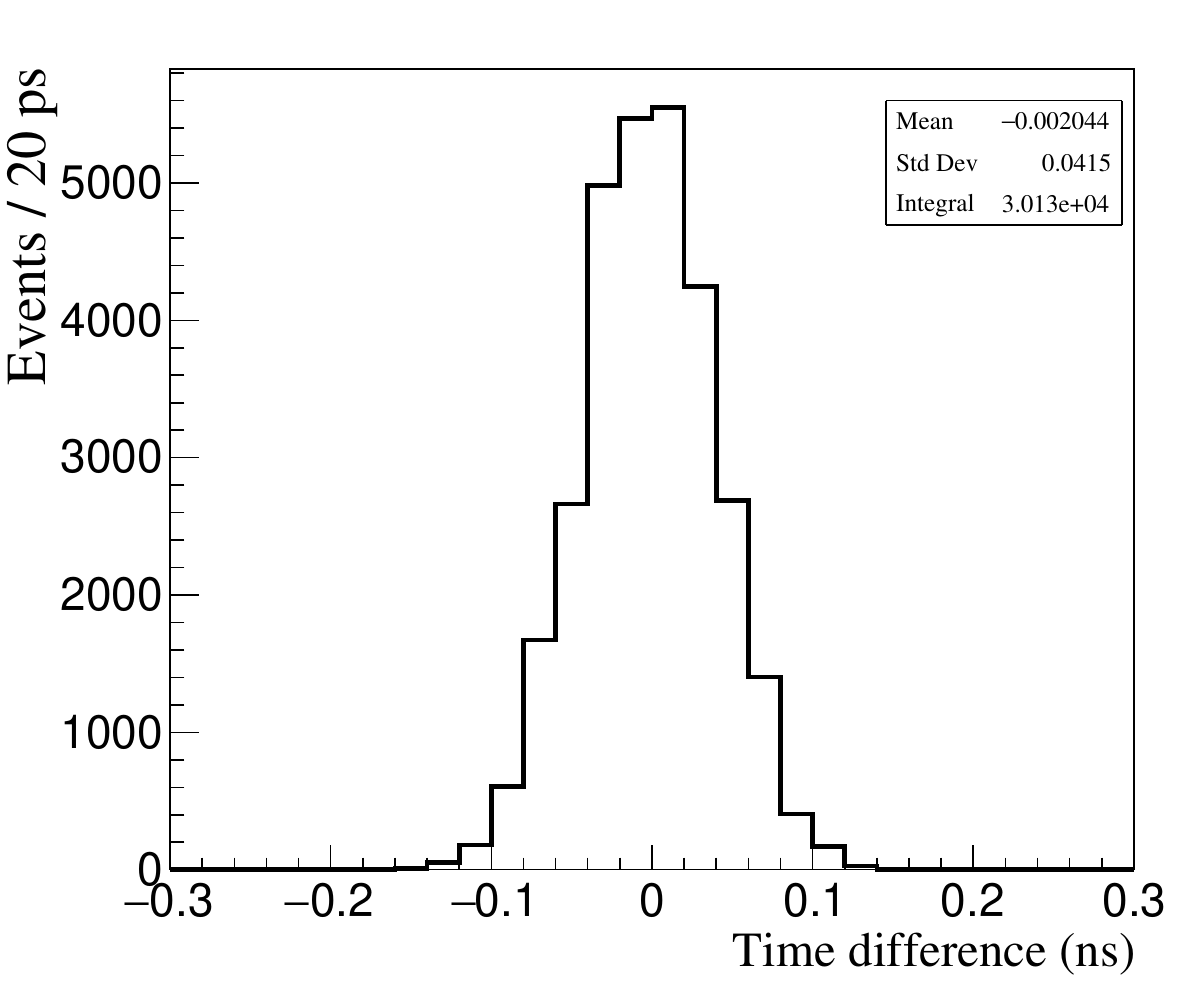}
  \caption{ The distribution of the time difference measured by two
    TDC channels after calibrations using identical SPE-like pulses, from a pulse
    generator that produces double-PE-like pulses which are then
    split equally.}
  \label{deltaT}
\end{figure*}

To evaluate the timing resolution of the TDC for comparator's output, we inject two identical
SPE-like pulses into two channels. The measured time difference is
shown in Figure~\ref{deltaT}. The RMS is much smaller than the transit
time spread (TTS) of the PMTs which is at the order of ns. Therefore, in
situ we expect to achieve a timing precision of O(ns) for PMTs. This
mainboard can also measure at most 24 ToT signals from SiPM arrays using the same
FPGA-based TDC technique.  Therefore, with SiPMs, even better timing
precision can be achieved. Previous study~\cite{Zhi:2024dmr} shows that
single photon time resolution with an array of $4\times4$ SiPMs is as
low as 300 ps FWHM, which is much smaller than the TTS of PMT. Notably, the IceCube-gen2
collaboration~\cite{IceCube-Gen2:2023rji} demonstrated that PMT digitized pulses can
be fitted with a smooth curve, achieving nanosecond-level time resolution.
In contrast, our TDC implementation enables direct hardware-level time
measurement. 

\section{Summary}
In summary, we presented a first fully custom-designed waveform and time
digitization mainboard prototype for the hDOM of the planned TRIDENT neutrino experiment in
the West Pacific Ocean. The prototype system demonstrates three key
capabilities: (1) simultaneous waveform digitization of up to 32 PMT
channels at 125 MS/s sampling rate; (2) single photoelectron
detection with clear separation from pedestal noise, while maintaining
good linearity up to 240 PEs; and (3) sub-nanosecond timing resolution for 32 PMT and
24 SiPM pulse leading edges, achieved through FPGA-based
time-to-digital converter implementation. These performance
characteristics make this mainboard suitable for the first hDOM prototype
string sea trial planned in Fall 2025.

\section{Acknowledgement}
This work has received support from the Ministry of Science and
Technology of China under Grant No.2022YFA1605500, the Office of
Science and Technology of the Shanghai Municipal Government under
Grant No.22JC1410100, and support from the Key Laboratory for Particle
Physics, Astrophysics and Cosmology, Ministry of Education.  We thank
Xin Xiang to provide the picosecond laser source for the linearity
evaluation.  We thank Hualin Mei to review and improve this paper.
On behalf of all authors, the corresponding author states that there
is no conflict of interest.

Data sets generated during the current study are available from the
corresponding author on reasonable request.

\bibliography{sn-bibliography}

\end{document}